# Classification of Epileptic EEG Signals by Wavelet based CFC

Amirmasoud Ahmadi,
Biomedical Engineering Department,
School of Electrical Engineering, Iran University of Science and Technology,
Tehran, Iran

Mahsa Behroozi,
Biomedical Engineering Department,
School of Electrical Engineering, Iran University of Science and Technology
Tehran, Iran

Vahid Shalchyan,
Biomedical Engineering Department,
School of Electrical Engineering, Iran University of Science and Technology
Tehran, Iran

Mohammad Reza Daliri
Biomedical Engineering Department
School of Electrical Engineering, Iran University of Science and Technology
Tehran, Iran

*Abstract*— Electroencephalogram, an influential equipment for analyzing human's activities and recognition of seizure attacks can play a crucial role in designing accurate systems which can distinguish ictal seizures from regular brain alertness, since it is the first step towards accomplishing a high accuracy computer aided diagnosis system (CAD). In this article a novel approach for classification of ictal signals with wavelet based cross frequency coupling (CFC) is suggested. After extracting features by wavelet based CFC, optimal features have been selected by t-test and quadratic discriminant analysis (QDA) have completed the Classification.

## I. INTRODUCTION

The common belief that seizure is a sign of epileptic brain disorder is not very accurate. Occurrence of seizure may take place regardless of circumstances or host's attributes [1]. Reports of WHO claims that the second plausible neurological disorder beneath stroke is epilepsy. Monitoring and diagnosing this considerable amount of afflicted patients is achievable by Electroencephalogram signals (EEG) [2,3]. Since this diagnosis requires physician's direct examination and results are not one and the same. Anticipation of epileptic seizure demands a method for automated computer aided diagnosis [4,5]. Time frequency domain [6,7], frequency domain [8-10] and time domain analysis [11] have been the basis of several feature extraction algorithms to detect seizure. After all, EEG signals are believed to be non-stationary. Between methods based on time frequency for feature extraction wavelet transforms are superior options due to their localization and reflection of time varying qualities of the data. Unprocessed EEG signals in conjunction with certain proper rules could be decompounded to precise subdivisions, consequently, noticeable amount of features are considered suchlike phase synchronization [12] effective correlation dimension [13], short term maximum Lyapunov exponents [14] accumulated energy [15] and dynamical similarity algorithm [16] to declare the existence of an epileptic seizure. A real time low power algorithm for classifying signals to detect seizures in ambulatory EEG was suggested by Patel et al [17]. Quadratic and linear discriminant analysis, support vector machine (SVM) and Mahalanobis discriminant analysis (MDA) classifiers have been examined in the aforementioned study on thirteen subjects.

In this article a new approach for ictal signals' classification based on wavelet is suggested after the extraction of optimal features from wavelet coefficients the signal has been classified by QDA and results claim that all cases have been designated correctly.

## II. MATERIALS AND METHODS

This study denotes a Wavelet based CFC method which firstly segregates the EEG signal into wavelet coefficients subsequently phase and amplitude of wavelet coefficients were computed with Hilbert transform. After ranking the optimal features by t-test, the classification procedure has been performed by QDA (Fig.1). Above mentioned cases are:

Case I: A vs. E (Healthy versus Seizure)

Case II: B vs. E (Healthy versus Ictal)

Case III: C vs. E (Hippocampal Interictal versus Seizure)

Case IV: D vs. E (Epileptogenic Interictal versus Seizure)

Case V: ABCD vs. E (Seizure-free versus Seizure)

### A. EEG Database of Epilepsy

The EEG database is accumulated from Germany epilepsy center, Bonn university hospital of Freiburg [18]. It consists of five subsets each containing 100 single channel EEG captured in international 10-20 electrode placement montage. In spite of the fact that C to E were captured intracranially, A and B have been recorded as extra cranial signals.

### B. Stationary Wavelet Transform (SWT)

The wavelet coefficients of SWT at all individual decomposition levels import the equal sample numbers same as the original signal. While DWT failed to face robustness and repeatability problems, SWT survived the obstacles [19].

### C. Feature Extraction based on Cross Frequency Coupling

Hilbert transform can indicate instantaneous phase and amplitude as follows: (M(n) is an analytic signal)

$$M(n) = m(n) + i\widetilde{m}(n) = A_m(n) \cdot e^{i\theta_m(n)} \qquad (1)$$

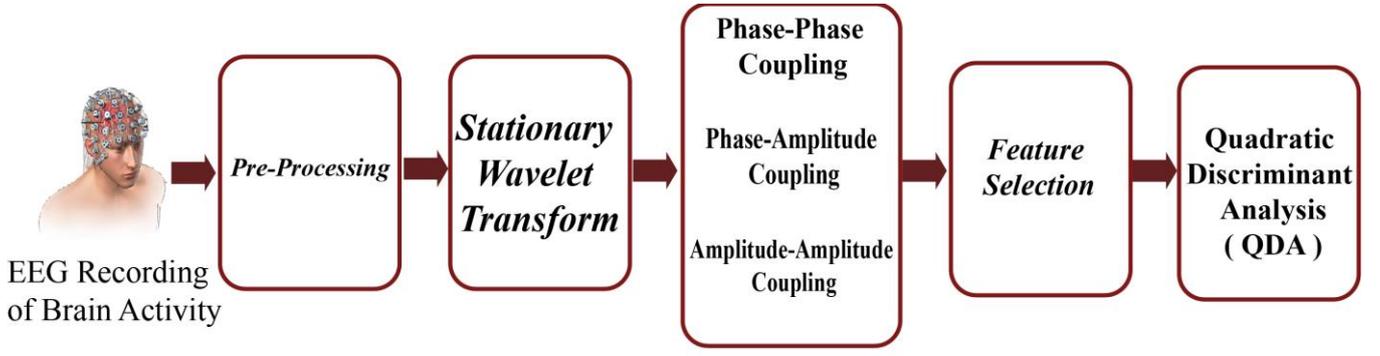

Fig. 1. Block diagram of proposed algorithm

Amplitude and instantaneous phase of the signals are defined as:

$$A_m(n) = \sqrt{m(n)^2 + \widetilde{m(n)}^2} \quad (2)$$

$$\theta_m(n) = \arctan(\frac{\widetilde{m(n)}}{m(n)}) \quad (3)$$

*1) Phase to Phase Coupling (PPC)*

Assuming that each x(n) and y(n) signals are oscillating in dissimilar wavelet coefficients, phase locking value could be estimated if $\Delta\theta(n) = T\theta_{m(n)} - R\theta_{y(n)}$ and signals are T:R synchronized. T=R=1 by reason of same references. PLV estimation could be done as follows and this amount can oscillate between zero (no coupling) and one (absolute coupling) [20]:

$$PLV = \left| \frac{1}{N} \sum_{n=1}^{N} e^{i\Delta\theta(n)} \right| \quad (4)$$

*2) Phase to Amplitude Coupling (PAC)*

At the time that Hilbert transform was done twice phase of an amplitude ($\phi_{amy}$) is measured to put to use in $P_{my}$ which is accepted as PLV for PPC [21].

$$P_{my} = \left| \frac{1}{N} \sum_{n=1}^{N} e^{i(\phi_y[n] - \phi_{amy}[n])} \right| \quad (5)$$

*3) Amplitude to Amplitude Coupling (AAC)*

Pearson correlation dissimilar to aforementioned methods employs correlation to represent coupling between +1 (absolute coupling) to -1 (no coupling) as follows:

$$Corr(M.Y) = \frac{COV(M.Y)}{\delta_m \delta_y} \quad (6)$$

*D. Feature Selection*

*1) T-test*

With regard to an excellent feature selection, for classification acceleration in computations should be done by expelling valueless features and retaining profitable ones. A statistical test, T-test, which operates based on arbitrating between quantities to check crucial divergent means have been applied in this research [22].

*E. Quadratic and linear discriminant analysis*

LDA applies linear hyper-planes in the character of decision surfaces to designate the input vector to different classes. The assumption of LDA as a classifier is that the non-identical Gaussian distributions are the basis of data generations in different classes. The training procedure can approximate the parameters with the fitting function. The train model looks for the class with the lowest misclassification cost presuming identicalness of covariance matrix (homoscedasticity). However, QDA is connected to LDA covariance matrices are not assumed to be the same and it form a quadratic decision boundary among classes [23].

## III. RESULTS AND DISCUSSION

In this article a novel approach for classification of ictal signals based on wavelet have been suggested. First, 'db4' mother wavelet SWT have been employed on all trials. Then, Hilbert transform were employed on wavelet coefficients so the phase and amplitude of each wavelet coefficient was calculated, finally PPC, PAC and AAC features were calculated.

In this research 10-fold cross validation was used in order to calculate the performance. 90% of trials were randomly selected as train data and the 10% residual were determined as the test data. The procedure was repeated 10 times and the average result was considered as the performance of systems. Extracted features of train data have been ranked by t-test and classification was performed by QDA. Optimal feature numbers were selected based on performance (Fig.2). In all cased trials were correctly classified (Table I). Choosing wavelet levels plays an important role in feature extraction. In this article based on system performance in different cases, 7 levels were experimented (Table II).

TABLE I. OBTAINED RESULTS CLASSIFYING EPILEPSY PATIENTS' EEG SIGNALS.

| Cases | Optimal number of features | Accuracy |
|---|---|---|
| Case I | 61 | 100 % |
| Case II | 15 | 100 % |
| Case III | 30 | 100% |
| Case IV | 40 | 100 % |
| Case V | 53 | 100 % |

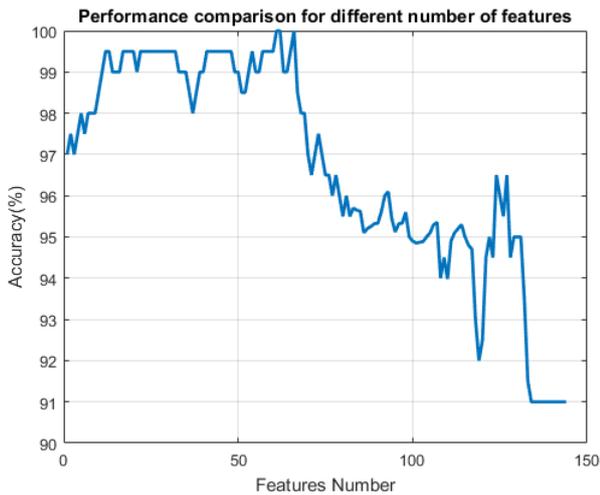

Fig. 2. Accuracy for different number of features

TABLE II. OBTAINED RESULTS OF DIFFERENT WAVELET DECOMPOSITION LEVELS

| | | Wavelet Decomposition Levels | | | | |
|---|---|---|---|---|---|---|
| | | 5 | 6 | 7 | 8 | 9 |
| CASES | A vs E | 100 | 100 | 100 | 100 | 100 |
| | B vs E | 100 | 100 | 100 | 100 | 100 |
| | C vs E | 99 | 99.5 | 100 | 100 | 100 |
| | D vs E | 99 | 99 | 100 | 100 | 99.5 |
| | ABCD vs E | 99.63 | 99.63 | 100 | 100 | 100 |

TABLE III. PROPOSED SYSTEM'S ACCURACY COMPARED WITH PREVIOUS SYSTEMS.

| Cases | Method | Accuracy |
|---|---|---|
| D Vs E | DWT based fuzzy approximate entropy + SVM [25] | 96 |
| | Simple random sampling + Clustering technique + LS-SVM [26] | 93.6 |
| | **Wavelet based CFC + QDA** | **100** |
| ABCD Vs E | Tunable-Q wavelet transform + Korsakov entropy + LS-SVM [27] | 97.75 |
| | CFC + t-test + Random Forest [28] | 99.87 |
| | **Wavelet based CFC + QDA** | **100** |

Selected optimal features have been evaluated from two aspects. Considering selected features in each case belongs to which wavelet coefficients and their participation ratio Fig.3 is presented. PAC, PPC and AAC were the aforementioned features. As an another point of view selected features were evaluated in order to find the most popular feature. Results declared that PPC hast the highest participation ratio among features.

In comparison with other suggested wavelet based methods, this novel approach claims a more significant performance, in all cases gaining 100 percent correct classification, tabulated in Table III.

## IV. CONCLUSION

In this research a novel method for classification of ictal EEG signals is suggested which has gained 100% correct classification performance. The innovation in feature extraction could completely clarify the differences between classes.

REFERENCES

[1] F. Mormann, T. Kreuz, C. Rieke, R. G. Andrzejak, A. Kraskov, P. David, *et al.*, "On the predictability of epileptic seizures," *Clinical neurophysiology,* vol. 116, pp. 569-587, 2005.
[2] W. H. Organization, "http://www. who. int/mediacentre/factsheets/fs340/en," *url> http://www. who. int/mediacentre/factsheets/fs241/en/</url,* 2014.
[3] R. B. Pachori and S. Patidar, "Epileptic seizure classification in EEG signals using second-order difference plot of intrinsic mode functions," *Computer methods and programs in biomedicine,* vol. 113, pp. 494-502, 2014.
[4] R. J. Martis, J. H. Tan, C. K. Chua, T. C. Loon, S. W. J. YEO, and L. Tong, "Epileptic EEG classification using nonlinear parameters on different frequency bands," *Journal of Mechanics in Medicine and Biology,* vol. 15, p. 1550040, 2015.
[5] L. Iasemidis, D.-S. Shiau, P. Pardalos, W. Chaovalitwongse, K. Narayanan, A. Prasad, *et al.*, "Long-term prospective on-line real-time

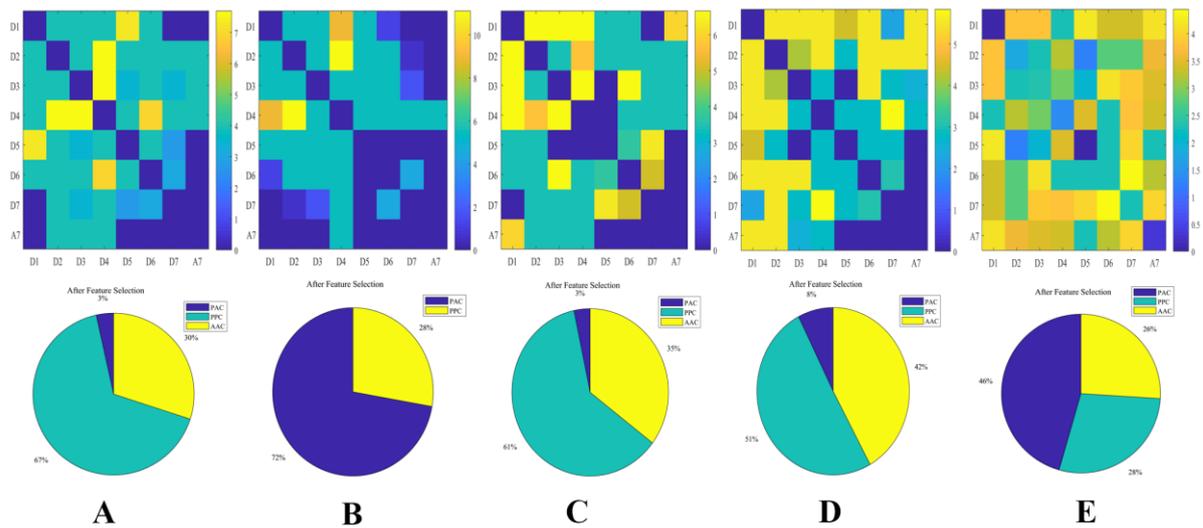

Fig. 3. **First row:** wavelet coefficients and their participation ratio. **Second row:** participation ratio among features

[6] seizure prediction," *Clinical Neurophysiology,* vol. 116, pp. 532-544, 2005.

[7] A. Ahmadi, V. Shalchyan, and M. R. Daliri, "A new method for epileptic seizure classification in EEG using adapted wavelet packets," in Electric Electronics, Computer Science, Biomedical Engineerings' Meeting (EBBT), 2017, 2017, pp. 1-4: IEEE.

[8] A. Ahmadi, S. Tafakori, V. Shalchyan and M. R. Daliri, "Epileptic seizure classification using novel entropy features applied on maximal overlap discrete wavelet packet transform of EEG signals," 2017 7th International Conference on Computer and Knowledge Engineering (ICCKE), Mashhad, 2017, pp. 390-395.

[9] J. Birjandtalab, M. B. Pouyan, and M. Nourani, ''Nonlinear dimension reduction for EEG-based epileptic seizure detection,'' in Proc. IEEEEMBS Int. Conf. Biomed. Health Inform., Feb. 2016, pp. 595–598.

[10] D. Cogan, J. Birjandtalab, M. Nourani, J. Harvey, and V. Nagaraddi, "Multi-biosignal analysis for epileptic seizure monitoring," International Journal of Neural Systems, 2016.

[11] J. Birjandtalab, M. B. Pouyan and M. Nourani, "An unsupervised subject identification technique using EEG signals," 2016 38th Annual International Conference of the IEEE Engineering in Medicine and Biology Society (EMBC), Orlando, FL, 2016, pp. 816-819.

[12] E. Tessy, P. P. M. Shanir and S. Manafuddin, "Time domain analysis of epileptic EEG for seizure detection," 2016 International Conference on Next Generation Intelligent Systems (ICNGIS), Kottayam, 2016, pp. 1-4.

[13] F. Mormann, K. Lehnertz, P. David, and C. E. Elger, "Mean phase coherence as a measure for phase synchronization and its application to the EEG of epilepsy patients," Physica D: Nonlinear Phenomena, vol. 144, pp. 358-369, 2000.

[14] K. Lehnertz and C. Elger, "Spatio-temporal dynamics of the primary epileptogenic area in temporal lobe epilepsy characterized by neuronal complexity loss," Electroencephalography and clinical Neurophysiology, vol. 95, pp. 108-117, 1995.

[15] L. D. Iasemidis, D.-S. Shiau, W. Chaovalitwongse, J. C. Sackellares, P. M. Pardalos, J. C. Principe, et al., "Adaptive epileptic seizure prediction system," IEEE transactions on biomedical engineering, vol. 50, pp. 616-627, 2003.

[16] B. Litt, R. Esteller, J. Echauz, M. D'Alessandro, R. Shor, T. Henry, et al., "Epileptic seizures may begin hours in advance of clinical onset: a report of five patients," Neuron, vol. 30, pp. 51-64, 2001.

[17] M. Le Van Quyen, J. Martinerie, M. Baulac, and F. Varela, "Anticipating epileptic seizures in real time by a non-linear analysis of similarity between EEG recordings," Neuroreport, vol. 10, pp. 2149-2155, 1999.

[18] K. Patel, C. Chem-Pin, S. Fau, C.J. Bleakley, Low power real-time seizure detection for ambulatory EEG, in: 3rd International Conference on Pervasive Computing Technologies for Healthcare, 2009, pp. 1–7.

[19] R. G. Andrzejak, K. Lehnertz, F. Mormann, C. Rieke, P. David, and C. E. Elger, "Indications of nonlinear deterministic and finite-dimensional structures in time series of brain electrical activity: Dependence on recording region and brain state," Physical Review E, vol. 64, p. 061907, 2001.

[20] Kayvanrad MH, McLeod AJ, Baxter JSH, et al. Stationary wavelet transform for under-sampled MRI reconstruction. Magn Reson Imag 2014; 32: 1353–1364.

[21] J. Weule, "Detection of n: m phase locking from noisy data: application to magnetoencephalography," Phys. Rev. Lett, vol. 81, pp. 3291-3294, 1998.

[22] B. Voytek, R. T. Canolty, A. Shestyuk, N. E. Crone, J. Parvizi, and R. T. Knight, "Shifts in gamma phase–amplitude coupling frequency from theta to alpha over posterior cortex during visual tasks," Frontiers in human neuroscience, vol. 4, 2010.

[23] M. Taghizadeh-Sarabi, M. R. Daliri, and K. S. Niksirat, "Decoding objects of basic categories from electroencephalographic signals using wavelet transform and support vector machines," Brain topography, vol. 28, pp. 33-46, 2015.

[24] Bishop, C.M. Pattern Recognition and Machine Learning. J. Electron. Imaging 2007, 16, 49901.

[25] H. Ocak, "Automatic detection of epileptic seizures in EEG using discrete wavelet transform and approximate entropy," Expert Systems with Applications, vol. 36, no. 2, pp. 2027-2036, 2009/03/01/ 2009.

[26] L. Guo, D. Rivero, J. Dorado, J. R. Rabunal, and A. Pazos, "Automatic epileptic seizure detection in EEGs based on line length feature and artificial neural networks," Journal of neuroscience methods, vol. 191, no. 1, pp. 101-109, 2010.

[27] Y. Kumar, M. Dewal, and R. Anand, "Epileptic seizure detection using DWT based fuzzy approximate entropy and support vector machine," Neurocomputing, vol. 133, pp. 271-279, 2014.

[28] A. Ahmadi, M. Behroozi, V. Shalchyan and M. R. Daliri, "Phase and amplitude coupling feature extraction and recognition of Ictal EEG using VMD," 2017 IEEE 4th International Conference on Knowledge-Based Engineering and Innovation (KBEI), Tehran, Iran, 2017, pp. 0526-0532.